\documentclass[nofootinbib,prl,superscriptaddress,a4paper,twocolumn]{revtex4-1}
\usepackage{geometry}
\usepackage[english]{babel}
\usepackage[latin1]{inputenc}
\usepackage{hyperref}
\usepackage{amsmath, amssymb, amsfonts, mathrsfs}
\usepackage{graphicx}
\usepackage{xspace}

\usepackage{braket}
\newcommand{\cnot}{{\small C}{\scriptsize NOT}\xspace}
\newcommand{\csum}{{\small C}{\scriptsize SUM}\xspace}
\newcommand{\cphase}{{\small C}{\scriptsize PHASE}\xspace}
\newcommand{\toffoli}{{\small T}{\scriptsize OFFOLI}\xspace}
\newcommand{\ccx}{{CC}{$\hat{X}$}\xspace}
\newcommand{\melvin}{{\small M}{\scriptsize ELVIN}\xspace}

\begin{document} 

\title{Computer-inspired concept for high-dimensional multipartite quantum gates}
\author{Xiaoqin Gao}
\affiliation{Faculty of Physics, University of Vienna, Austria.}
\affiliation{Institute for Quantum Optics and Quantum Information (IQOQI) Vienna, Austrian Academy of Sciences, Austria.}
\affiliation{National Mobile Communications Research Laboratory \& Quantum Information Research Center, Southeast University, Nanjing, China.}
\author{Manuel Erhard}
\affiliation{Faculty of Physics, University of Vienna, Austria.}
\affiliation{Institute for Quantum Optics and Quantum Information (IQOQI) Vienna, Austrian Academy of Sciences, Austria.}
\author{Anton Zeilinger}
\affiliation{Faculty of Physics, University of Vienna, Austria.}
\affiliation{Institute for Quantum Optics and Quantum Information (IQOQI) Vienna, Austrian Academy of Sciences, Austria.}
\author{Mario Krenn}
\email{mario.krenn@univie.ac.at}
\affiliation{Faculty of Physics, University of Vienna, Austria.}
\affiliation{Institute for Quantum Optics and Quantum Information (IQOQI) Vienna, Austrian Academy of Sciences, Austria.}
\affiliation{Department of Chemistry \& Computer Science, University of Toronto, Canada.}
\affiliation{Vector Institute for Artificial Intelligence, Toronto, Canada.}

\begin{abstract}
An open question in quantum optics is how to manipulate and control complex quantum states in an experimentally feasible way. Here we present concepts for transformations of high-dimensional multi-photonic quantum systems. The proposals rely on two new ideas: (I) a novel high-dimensional quantum non-demolition measurement, (II) the encoding and decoding of the entire quantum transformation in an ancillary state for sharing the necessary quantum information between the involved parties. Many solutions can readily be performed in laboratories around the world, and identify important pathways for experimental research in the near future. The concept has been found using the computer algorithm \melvin for designing computer-inspired quantum experiments. This demonstrates that computer algorithms can inspire new ideas in science, which is a widely unexplored potential.
\end{abstract}

\maketitle

\section{Introduction}
One collective goal of quantum optics research is to find ways for controlling complex quantum systems, both for investigating fundamental questions of quantum mechanics and for potential applications in quantum technology \cite{pan2012multiphoton, flamini2018photonic}. 

The complexity of a quantum system increases both with the number of involves parts, as well as the number of dimensions of these individual parts. For single photonic quantum systems, it is known for 25 years how to perform arbitrary unitary transformations \cite{reck1994experimental}, which has since become a foundation for integrated photonics \cite{schaeff2015experimental,smania2016experimental, wang2018multidimensional, paesani2019generation}. Also in other degrees-of-freedom of photons, single quDit quantum gates have been well understood, for example using discretized time steps \cite{kues2017chip} or spatial modes of photons \cite{zhang2016engineering, babazadeh2017high,gao2019arbitrary,brandt2019high}, and high-dimensional multi-degree-of-freedom operations on single photons \cite{imany2019high}.

Multi-photon operations are more intricated, as photons do not interact with each other. To overcome this difficulty and perform an effective interaction between two photons, ancillary states are used to herald probabilistic transformations, such as controlled-NOT (\cnot) gates \cite{gasparoni2004realization, zeuner2018integrated}. The quality of these transformations has immensely increased, enabling on-chip demonstrations of arbitrary two-dimensional two-photon gates, as well as theoretical concepts for arbitrary photonic quBit transformations \cite{qiang2018large}. Summing up, the special cases of multiphotonic qubit transformations, and single-photonic arbitary high-dimensional transformations are well understood. However, the general case of transformations of $n$ photons in $d$ dimensions is still open. 

Here we show blueprints for experimental realizations of arbitrary multidimensional multiphotonic transformations. The concept is based on the encoding the essence of quantum transformations in an ancilla state, which mediates the neccessary quantum information between the involved photons. This is made possible by the introduction of a new quantum nondemolition measurement and the exploitation of a genuine high-dimensional interferometer. Several of our experimental proposals are feasible with state-of-the-art technology, whole others show an important route that needs to be investigated in the future.

\begin{figure*}[!t]
\centering
\includegraphics[width=0.85\textwidth]{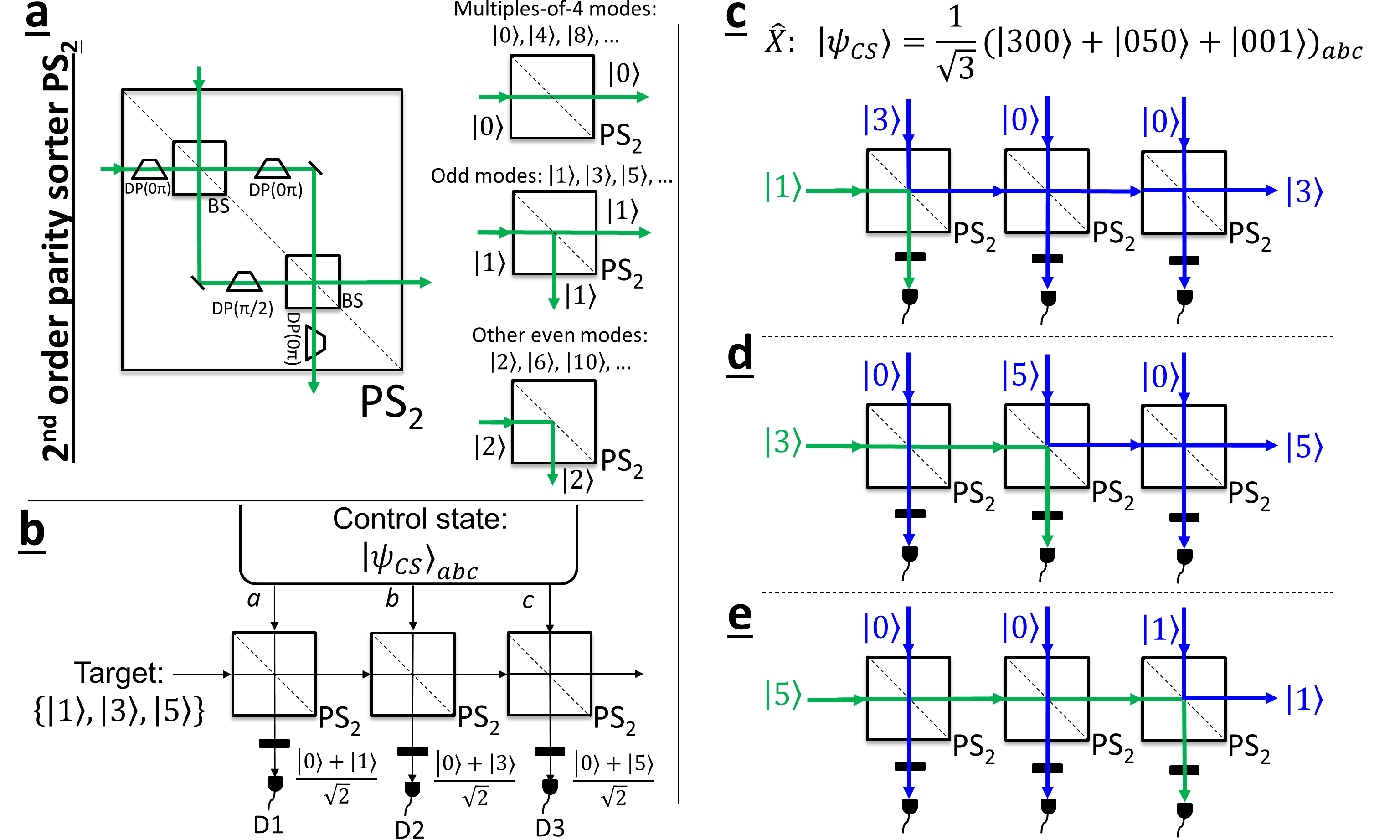}
\caption{An ancilla state based setup for three-dimensional control gates $\hat{X}$, $\hat{X}^2$ and $\hat{X}^3$. \textbf{\underline{a:}} A genuinely three-dimensional action of a second-order parity sorter (PS$_2$). On the left is the physical implementation as an interferometer with Dove prisms (DP) which introduces a reflection and a mode-dependent phase of of $\alpha=0$ or $\alpha=\ell\pi/2$, introduced by Leach et al.\cite{leach2002measuring}. Right, the abstract representation of the element. It can separate deterministically photons with mode $\ket{0}$ (transmitted) from photons with mode number $\ket{2}$ (reflected). Crucially for our requirements is that photons with odd mode numbers are randomly separated into two output paths, in analogy to a conventional beam splitter. Thus, the PS$_2$ is a 2-input, 2-output element, which can perform entirely different transformations on three orthogonal modes. \textbf{\underline{b:}} The state of a 3-photon state $\ket{\psi_{CS} }$ determines the transformation of the target photon. The target photon is overlapped with the 3 ancilla photons at a PS$_2$ each. One output mode is connected to a detector, which heralds a correct transformation. If all detectors see a photon, the transformation was successful and the output photon has the correct state. Importantly, PS$_2$ transmits modes $\ket{0}$ and probabilistically splits odd mode numbers. The state is encoded in the odd number space \{$\ket{1}$,$\ket{3}$,$\ket{5}$\}. The desired transformation can be achieved by adjusting the ancilla state. \textbf{\underline{c-e:}} We show the $\hat{X}$ transformation in detail. The ancilla state for the $\hat{X}$ transformation is $\ket{\psi_{CS} }=\frac{1}{\sqrt{3}}\left(\ket{3,0,0}+\ket{0,5,0}+\ket{0,0,1}\right)$. If the target photon is $\ket{1}$, the only way to have all detectors see a photon (after the filter) is to have the photon in path $a$ beeing $\ket{3}$ and leaving to the output port, and the photons in path $b$ and $c$ both being in state $\ket{0}$ and going to detectors D2 and D3. This happens with a probability of p=$0.5^4$, because odd modes need to reflect in a specific way four times. The other two terms of $\ket{\psi_{CS} }$ will not be able to produce all three detectors D1-D3 clicking. The state $\ket{0,5,0}$ will not be able to create a click in detector D2, and $\ket{0,0,1}$ will not produce a click in either D1 or D3. Therefore, the pattern in the image is the only possible combination. Similar reasoning leads to the conclusions for the target photons being in $\ket{3}$ and $\ket{5}$. Exactly the same logic holds for the other two types of transformations, $\hat{X}^2$ with $\ket{\psi_{CS} }=\frac{1}{\sqrt{3}}\left(\ket{5,0,0}+\ket{0,1,0}+\ket{0,0,3}\right)$ and $\hat{X}^3$ with $\ket{\psi_{CS}}=\ket{0,0,0}$.}
\label{fig:TargetPart}
\end{figure*}
\textit{Inspirations from computers -- } We have identified the initial concepts presented here via computer-designed experiments using the algorithm \melvin \cite{krenn2016automated}. To do that, we had to formulate the question of effective photon-photon interaction in the most general and precise way. The total search space of quantum optical setups here corresponds to roughly $10^{30}-10^{40}$ possibilities. With efficient exclusion principles\footnote{Exclusion principles determine when setups should not be investigated in detail. For example, an experimental configuration where the quantum information of the control photon cannot reach the target photon can be excluded without calculation. Many cases can be excluded, thus leading to much more efficient search procedures.}, we were able to significantly reduce the number of setups that needed to be calculated. A total of roughly 150.000 CPU hours has then finally uncovered the seed of ideas we present in this manuscript. The human scientists in our team were subsequently able to generalize the computer-inspired ideas to the form presented here.

\section{High-dimensional control operations}
The simplest non-trivial case of a multi-photonic transformation is a 2-dimensional \cnot gate, where the state of the a target photon changes depending on the state of a control photon. The four transformations are written as 
\begin{align}
	\textnormal{\cnot}\ket{0,0}=\ket{0,0}&,& \textnormal{\cnot}\ket{0,1}=\ket{0,1}, \nonumber\\
	\textnormal{\cnot}\ket{1,0}=\ket{1,1}&,& \textnormal{\cnot}\ket{1,1}=\ket{1,0}. 
  \label{2d-cnot}
\end{align}
A more compact way is
\begin{align}
	\textnormal{\cnot}\ket{c,t}=\ket{c}\hat{X}^c\ket{t}=\ket{c,(c+t)\%2}
  \label{2d-cnot2}
\end{align} 
where \% stands for the modulo operation, and $\hat{X}$ stands for the Pauli-X operation (with $\hat{X}\ket{0}=\ket{1},\hat{X}\ket{1}=\ket{0}$, or more compact, $\hat{X}\ket{n}=\ket{(n+1)\%2}$). Thus, one can think about the \cnot as an $\hat{X}$ operation applied $c$ times on the target photon.  Crucially, the \cnot operates coherently on superpositions of terms in eq.(\ref{2d-cnot}), which distinguishes it from classical operations, and enables its usage in quantum applications.

We generalize the concept to high-dimensional systems \cite{gottesman1998fault, daboul2003quantum, bocharov2017factoring}. A high-dimensional generalisation of the \cnot is a controlled-$\hat{X}$, $\hat{CX}$, which acts as
\begin{align}
	\hat{\textnormal{CX}}\ket{c,t}=\ket{c}\hat{X}_d^c\ket{t}=\ket{c,(c+t)\%d},
  \label{d-cx}
\end{align}
with the high-dimensional Pauli-$\hat{X}$ gate acting as $\hat{X}\ket{n}=\ket{(n+1)\%d}$. Informally, we increase the value of the target photon by the value of the control photon (modulo $d$). This can also be considered as a \csum gate. Generalisations of other important quantum gates are controlled-controlled-$\hat{X}$ gates (\ccx, which is a generalisation of the important three-qubit Toffoli gate) which acts on two control photons and one target as $\textnormal{\ccx}\ket{c_1,c_2,t}=\ket{c_1,c_2}\hat{X}^{c_1\cdot c_2}_d\ket{t}=\ket{c_1,c_2,(t+c_1\cdot c_2)\%d}$, and \cphase which acts as $\textnormal{\cphase}\ket{c_1,c_2}=w^{c_1\cdot c_2}\ket{c_1,c_2}$ (which are essential components in the generation of graph states, that are used in measurement-based quantum computing \cite{raussendorf2001one}). 

\section{Experimental concepts}
Here we present the experimental concepts to perform high-dimensional multi-photonic transformations. We consider the orbital-angular momentum of photons as the discrete degree-of-freedom \cite{rubinsztein2016roadmap, erhard2018twisted}, but our concept is general enough that it can be translated to any other discrete high-dimensional system, such as path-encoding or time-encoding, or potentially even beyond photons.

First we explain how a well-studied 2-input 2-output optical element \cite{leach2002measuring} can be used in a genuine 3-dimensional way. This element is crucial for what comes next, a step by step construction of a 3-dimensional generalisation of a \cnot: We show how a three-photon ancilla state can effectively transform a 3-dimensional target photon. Then we introduce a new experimental non-destructive measurement scheme to extract the quantum information of a 3-dimensional control photon. The combination of these two structures leads to a full 3-dimensional $C\hat{X}$ gate. Finally we show experimental methods to generate other, more complex multi-photonic transformations. 

\textit{Genuine high-dimensional 2-input 2-output element} 
Conventionally, 2-input 2-output elements perform either the same action on all incoming modes (such as the beam splitter, which transmits or reflects an incoming photon purely by chance), or performs two different actions on two classes of modes (such as the polarizing beam splitter, which transmits horizontally polarised photons, and reflects vertically polarized ones). 
\begin{figure}[b]
\centering
\includegraphics[width=0.5\textwidth]{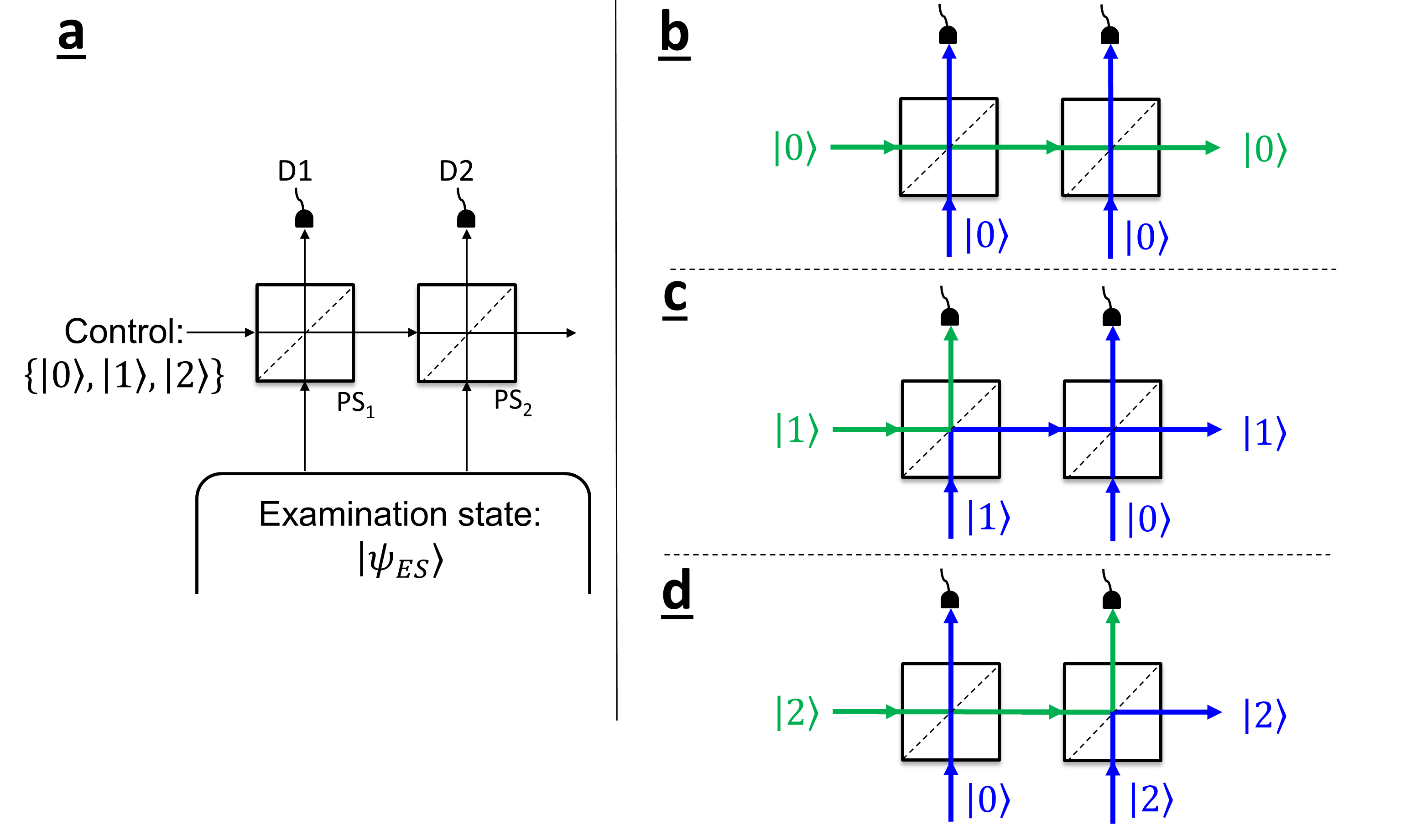}
\caption{Examining the control photon without destruction: Quantum nondemolition measurement. The goal is to extract information of the control photon without destroying it. \textbf{\underline{a:}} We combine the control photon with a two-photon examination state $\ket{\psi_{ES}}$ at a PS$_1$ (which transmits even modes and recflects odd modes) and a PS$_2$ (as described in Fig. \ref{fig:TargetPart}a). \textbf{\underline{b-d:}} For each of the possible control states $\ket{0}$, $\ket{1}$ and $\ket{2}$, there is only one combination which produces clicks in the detectors D1 and D2. If both detectors register a photon, the output photon has the correct mode number. Thus we can probe the control photon with three different ancilla states and we know its state, heralded by the coincident detections of a photon in both detectors.}
\label{fig:ControlledPart}
\end{figure}

\begin{figure*}[!t]
\centering
\includegraphics[width=0.95\textwidth]{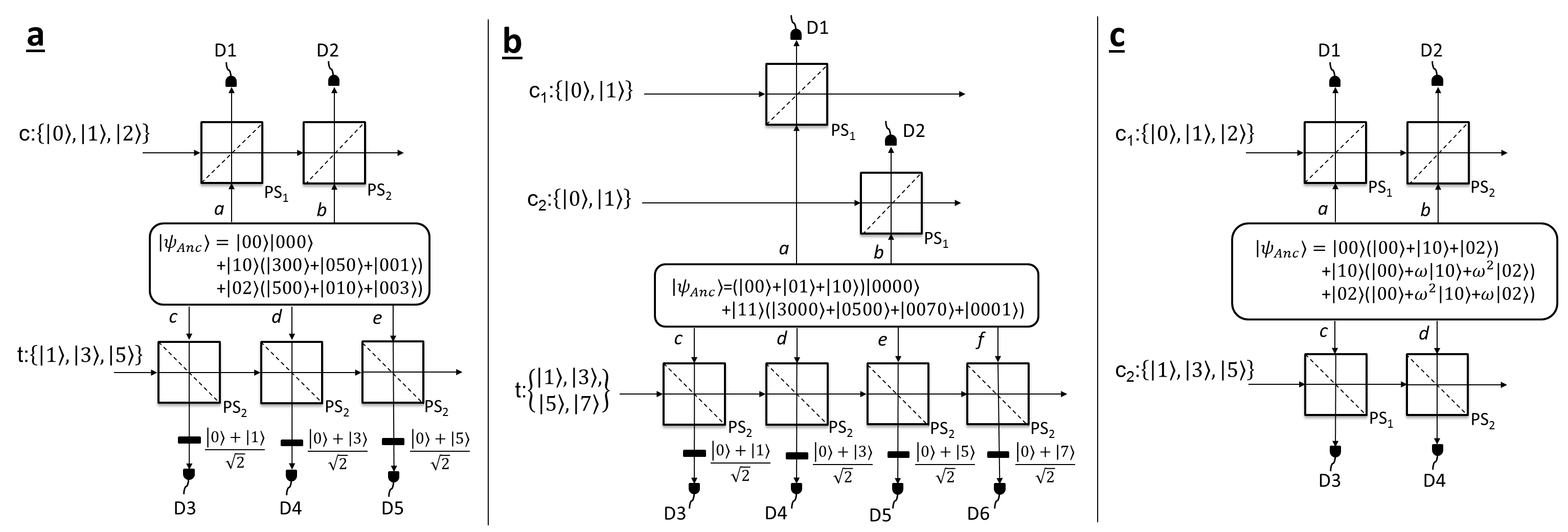}
\caption{Three high-dimensional multi-photonic quantum transformations. \textbf{\underline{a:}} A high-dimensional two-particle three-dimensional controlled-$\hat{X}_3$ gate, which can be described as $\hat{T}\ket{c,t}=\ket{c}\hat{X}^c\ket{t}=\ket{c,(c+t)\%3}$. It is a combination of the elements in Fig. \ref{fig:TargetPart} (for the target photon) and Fig. \ref{fig:ControlledPart} (for the control photon). If all detectors D$_1$-D$_5$ click simultaneously, then the transformation was successful and the two photons are emitted in the correct state. \textbf{\underline{b:}} A controlled-controlled-$\hat{X}_4$ quantum gate, with two control photons being in a two-dimensional state while the target photon is in a four-dimensional state. The explicit transformation is $\hat{T}\ket{c_1,c_2,t}=\ket{c_1,c_2}\hat{X}^{c_1\cdot c_2}\ket{t}=\ket{c_1,c_2,(t+c_1\cdot c_2)\%4}$. \textbf{\underline{c:}} A 3-dimensional controlled phase operation sees twice the method of Fig. \ref{fig:ControlledPart} (as the mode does not change, only the phases). The transformation is $\hat{T}\ket{c_1,c_2}=\omega^{c_1\cdot c_2}\ket{c_1,c_2}$.}  
\label{fig:CompleteExamples}
\end{figure*}

A key idea that enables all what follows is the understanding that a well-established experimental element can actually perform three different transformations, see Fig. \ref{fig:TargetPart}a. A second-order parity sorter will transmit modes $\ket{0}$, $\ket{4}$, $\ket{8}$, \dots and reflect $\ket{2}$, $\ket{6}$, $\ket{10}$, \dots. Interestingly, for odd modes it performs another transformation -- it splits the wavefunction by chance into an reflected and transmitted beam (just as a beam splitter). This is important because in this way, we can mix deterministic and probabilistic operations at a single element.  

\textit{Controlling a photon by other photons} --
Now we show how the state of a 3-dimensional photon can be controlled with three the states of other photons. The correct transformation is heralded by the simultaneous clicks of three detectors D1-D3, see Fig. \ref{fig:TargetPart}b. The three photons from a controlling state $\ket{\psi_{CS}}$ are combined with a target photon at three $PS_2$ (explained in Fig. \ref{fig:TargetPart}a). Infront of every detector is a mode filter, which projects the photons into the state $\frac{1}{\sqrt{2}}\left(\ket{0}+\ket{1}/\ket{3}/\ket{5} \right)$. A photon in state $\ket{0}$ is deterministically transmitted to the detectors. An odd photon splits probabilistically at the PS$_2$. In order to produce all detectors fire (and a photon at the output), the state $\ket{\psi_{CS}}$ is prepared in such a way that, the target input photon has to go to one of the detectors and the output photon is replaced by one from $\ket{\psi_{CS}}$. This leads to one single possible configuration how the detectors D1-D3 can fire, which are shown in Fig.\ref{fig:TargetPart}b. All other configurations lead to at least one detector not fireing, or no photon leaving the setup. By changing the $\ket{\psi_{CS}}$, we can perform various other manipulations of the target photon. The three examples of $\ket{\psi_{CS}}$ in Fig.\ref{fig:TargetPart}c-e will lead to a $\hat{X}$ transformation, just controlled by the three photons of $\ket{\psi_{CS}}$. 

\textit{Quantum Nondemolition Measurement and a 3-dimensional C$\hat{X}$} -- Our goal is to generate, as a first example, a 3-dimensional $C\hat{X}$ ($C\hat{X}\ket{c,t}=\ket{c}\hat{X}^c\ket{t}$), using an ancilla state that mediates the information from the control photon to the target photon. Above we have already seen how $\ket{\psi_{CS}}$ can control a target photon. Of course, later $\ket{\psi_{CS}}$ will become part of the mediating ancilla state. The remaining question now is how to extract information from the control photon (in an examination state $\ket{\psi_{ES}}$) without destroying its quantum information. We employ a similar idea as before, using two ancilla photons and two detectors -- the quantum information is correctly extracted if the two detectors fire simultaneously. In our example, $\ket{\psi_{ES}}$ can be one out of three quantum states. Each of them makes detectors D1 and D2 click for a different input state. For example, if $\ket{\psi_{ES}}=\ket{0,0}$, the two detectors will click and a photon will output the setup only if the input photon was in state $\ket{0}$. In all other cases, either no photon exits the setup (which can be considered as loss) or not all detectors fire simultaneously. However, if they click, we know the photon was in the state $\ket{0}$ without destroying its quantum information.

Now we have the two important functionalities: We can extract the quantum information from the input state without destroying it, and we can control the transformation applied to a target state using ancillary photons. Finally we can combine these two ingredients. We use an entangled state that combines $\ket{\psi_{ES}}$ and $\ket{\psi_{CS}}$ into $\ket{\psi_{Anc}}$, as shown in Fig. \ref{fig:CompleteExamples}a. For example, if the input state is $\ket{1}$, only $\psi_{ES}=\ket{1,0}$ can lead to clicks in detectors D1 and D2, thus $\ket{\psi_{Anc}}$ collapses into $\ket{\psi_{Anc}}_{1}=\frac{1}{\sqrt{3}}\left(\ket{3,0,0}+\ket{0,5,0}+\ket{0,0,1} \right)$. This state introduces an $\hat{X}$ transformation at the target photon, exactly as described in Fig. \ref{fig:TargetPart}. Therefore, a click in detectors D1-D5 heralds a successful 3-dimensional controlled-$\hat{X}$ transformation.

\textit{General multiphotonic high-dimensional transformations} -- We can apply the same idea to more complex transformations, such as controlled-controlled-$\hat{X}$ in Fig. \ref{fig:CompleteExamples}b or a controlled-Phase gate in Fig. \ref{fig:CompleteExamples}c. In both cases, we apply the same concepts as shown in Figs. \ref{fig:TargetPart} and \ref{fig:ControlledPart}. More general transformation, which are not simple control operations (an example is $\hat{T}\ket{c_1,c_2}=\ket{(c_1+c_2)\%d,(2c_1+c_2)\%d}$) can be generated in a very similar way, which we show in the SI.

One important remaining question is how to create the ancillary state experimentally. The last five years has seen a plethora of high-dimensional multiphotonic experiments \cite{wang2015quantum, malik2016multi, hiesmayr2016observation, zhang2017simultaneous, erhard2018experimental}, indicating that large classes of quantum states are accessible.  Recently, \textit{entanglement by path identity} has been proposed \cite{krenn2017entanglement} which is a conceptually very efficient method to produce high-dimensional, multi-photonic entanglement. A map from entangled states to graph theory allows to analyse the generation of these states in a systematic way, see in particular \cite{krenn2017quantum, gu2019quantum} and the first experimental demonstration \cite{kysela2019experimental}. In addition, high-dimensional Bell-state measurements \cite{luo2019quantum, zhang2019arbitrary} can be used to entangle separated quantum states and quantum teleportation can be used to probe the photon number in various paths \cite{wang2015quantum}. We show in the SI how to create several of these states using well-known experimental concepts.

\section{Conclusion and Outlook}
We have presented general multi-photon high-dimensional transformations, which solely rely on known experimental techniques. Among them are high-dimensional generalisations of the crucial \cnot, \toffoli or \cphase gates. Several of these concepts can readily be implemented in laboratories around the world.

The experimental configurations presented here are probabilistic, and require a number of additional ancillary photons to be performed correctly. This is expected, as the same holds true for the well-studied two-dimensional case. In order to improve the practicality of these setups, it would be interesting to study how (if possible) to reduce the number of ancillary photons and increase the success probability of the gates.

For experimental implementation, high stability of the experimental setups will be necessary. This can be done in bulk optics (such as shown here \cite{luo2019quantum}), or by implementing these methods into integrated circuits \cite{schaeff2015experimental, wang2018multidimensional}, ideally with the possibility of having photon-pair sources on-chip \cite{silverstone2014chip, jin2014chip, krapick2016chip,atzeni2018integrated}, to produce ancilla states in a stable manner. Therefore, it will be interesting to translate the concepts presented here to the path-encoding degree-of-freedom.

The concepts for these gates have been discovered using computer-designed quantum experiments, specifically a highly efficient version of the algorithm \melvin\cite{krenn2016automated}. Several other automated algorithms have been generated recently for the design of novel quantum-optical experiments \cite{melnikov2018active, o2019hybrid, wallnofer2019machine}. Our result indicates the possibility that computers can be used in a widely unexplored way, namely for inspirations of human scientists. 

\section*{Acknowledgements}
The authors thank Borivoje Dakic, Giuseppe Vitagliano, Robert Fickler, Paul Appel, Marcus Huber, Nicolai Friis, Fabien Clivaz, Jaroslav Kysela and Yelena Guryanova for helpful discussions. XQG thanks Bin Sheng and Zaichen Zhang for support. This work was supported by the Austrian Academy of Sciences ({\"O}AW), and the Austrian Science Fund (FWF) with SFB F40 (FOQUS). XQG acknowledges support from the National Natural Science Foundation of China (No. 61571105) and the China Scholarship Council (CSC). MK acknowledges support from FWF via the Erwin Schr\"odinger fellowship No. J4309.

\bibliographystyle{unsrt}
\bibliography{refs}

\clearpage
\newpage

\begin{center}
\textbf{\Large Supplementary Information}
\end{center}

\subsection{More general, non-controlled high-dimensional multi-photonic operations}
We find concepts for more general multi-photonic transformations which can not be interpreted as a controlled-operation, where one photon controls the operation of the other one. As an example, consider
\begin{figure}[b]
\centering
\includegraphics[width=0.5\textwidth]{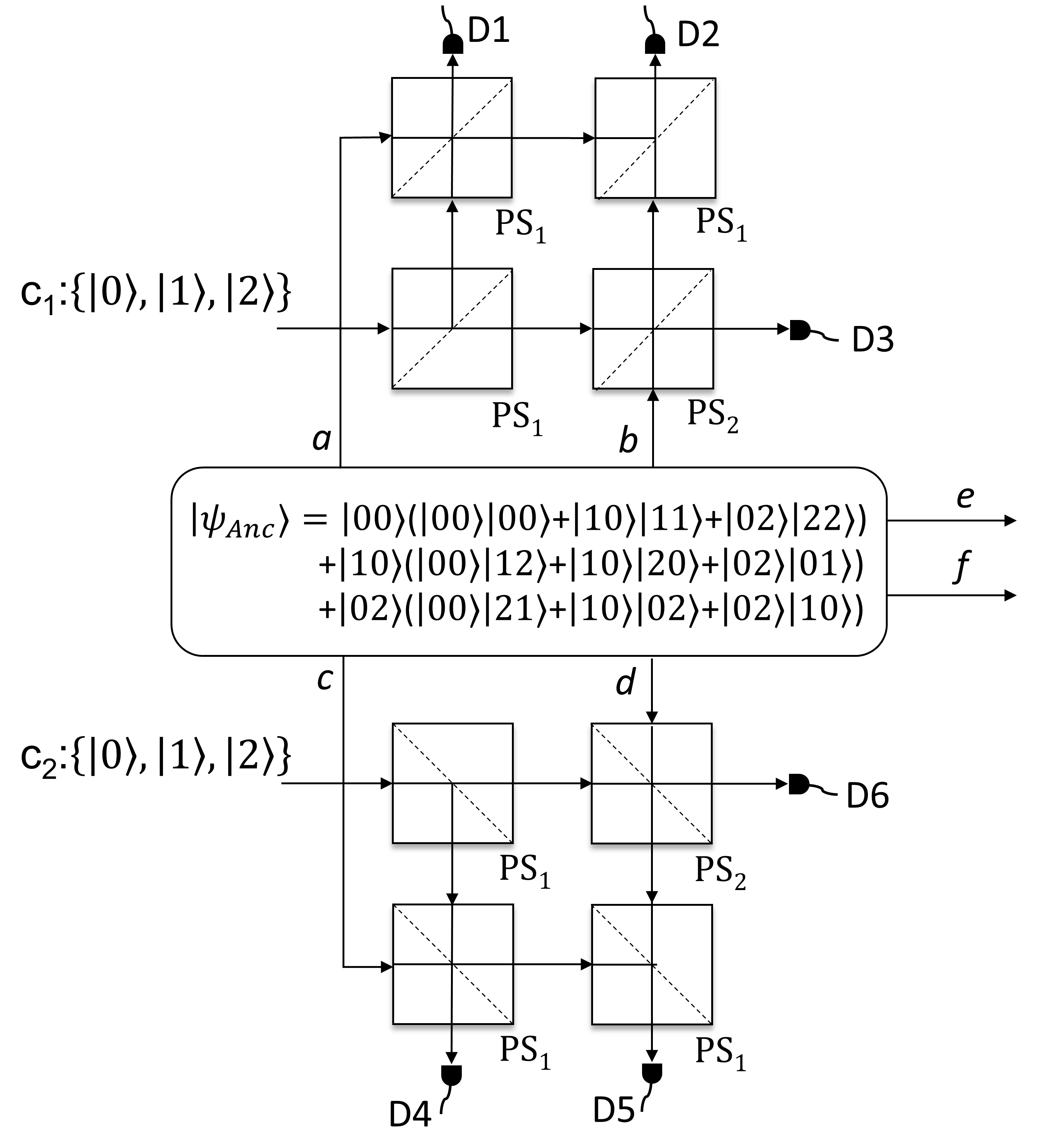}
\caption{Two photons which encode quantum information in three levels each, ($c_1, c_2 \in \ket{0},\ket{1},\ket{2}$) are transformed according to eq.\ref{generalTrafo}. A six-particle ancilla state encodes the transformation information. A symmetric combination of PS$_1$ and PS$_2$ are used to extract the quantum information of the two photons and transform them accordingly. The quantum information is transmitted via the highly entangled $\ket{\psi_{Anc}}$ state.}
\label{fig:SIgenTrafo}
\end{figure}
\begin{align}
	\hat{T}\ket{c_1,c_2}=\ket{(c_1+c_2)\%d,(2c_1+c_2)\%d}.
  \label{generalTrafo}
\end{align}
If $d=3$, this leads to the following transformation:
\begin{align}
	\hat{T}\ket{0,0}&=\ket{0,0}, \hat{T}\ket{0,1}=\ket{1,1}, \hat{T}\ket{0,2}=\ket{2,2}\nonumber\\
	\hat{T}\ket{1,0}&=\ket{1,2}, \hat{T}\ket{1,1}=\ket{2,0}, \hat{T}\ket{1,2}=\ket{0,1}\nonumber\\
  \hat{T}\ket{2,0}&=\ket{2,1}, \hat{T}\ket{2,1}=\ket{0,2}, \hat{T}\ket{2,2}=\ket{1,0}.
  \label{generalTrafo}
\end{align}
Such a transformation can be created Fig.\ref{fig:SIgenTrafo}. The transformation is encoded into a six-photon ancilla state. The detection of a photon in each of the six detectors D1-D6 herald a correct transformation.

\subsection{Creating the ancillary state for 2-dimensional Control and d-dimensional Target}
The ancillary are complex high-dimensional, multipartite entangled state and it is not clear per se how to create them experimentally. We show, for an important class of transformations how they can be created, namely for transformations where the control photon resides in two dimensions, while the target photon is encoded in a $d$-dimensional space.

\begin{figure}[b]
\centering
\includegraphics[width=0.4\textwidth]{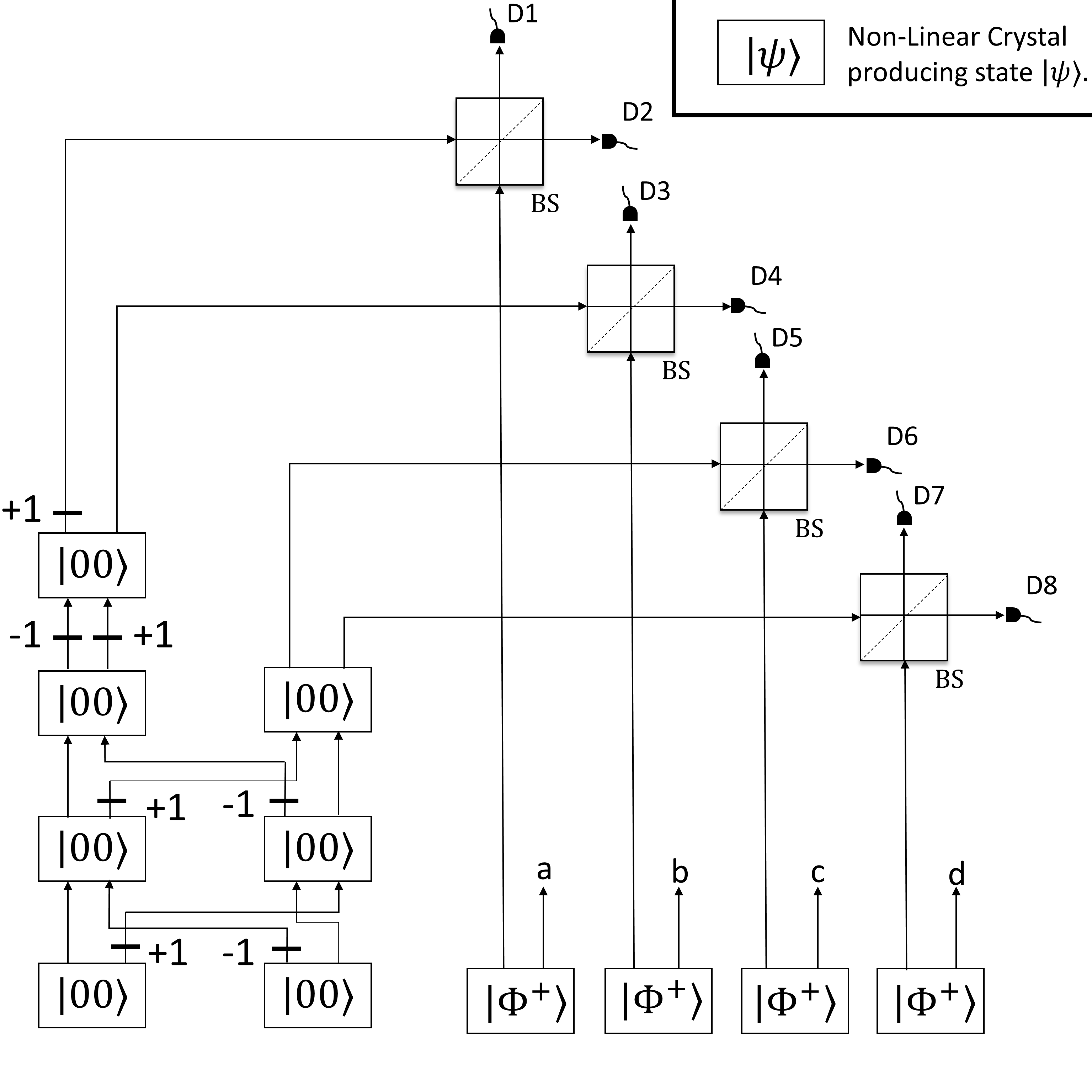}
\caption{An experimental setup for a 2-dimensional control, 3-dimensional target ancillary state in eq.(\ref{statec2t3_W}). Simultaneous clicks of detectors D1-D8 indicate the generation of the correct ancilla state in path $a$-$d$. It is based on \textit{Entanglement by Path Identity} \cite{krenn2017entanglement} and a multi-photon filter using Bell-State measurements introduced in \cite{wang2015quantum}. }
\label{fig:SIW}
\end{figure}

The ancillary state for such a transformation in the paths $a$,$b$,$c$,$d$ is
\begin{align}
	\ket{\psi}=\frac{1}{2}(\ket{0,0,0,0}&+\ket{1}(\ket{3,0,0}+\nonumber\\
  &+\ket{0,5,0}+\ket{0,0,1}))_{abcd}
  \label{statec2t3}
\end{align}
which is equivalent (up to local transformations) to
\begin{align}
	\ket{\psi}=\frac{1}{2}(\ket{1,0,0,0}&+\ket{0,1,0,0}+\nonumber\\
  &+\ket{0,0,1,0}+\ket{0,0,0,1})_{abcd}.
  \label{statec2t3_W}
\end{align}
Equation (\ref{statec2t3_W}) is a W-state \cite{zeilinger1992higher}. It can be experimentally created using \textit{Entanglement by Path Identity} \cite{krenn2017entanglement}. Importantly the state must not be post-selected, which means that each path has exactly one photon, not multiple photons. This can be guaranteed by the exploitation of Bell-state measurements (as introduced in \cite{wang2015quantum}). The full experimental setup is shown in Fig. \ref{fig:SIW}. 

A generalisation of the setup to transformation of 2-dimensional control, $d$-dimensional target requires ($d$+1)-photonic W state. The experimental generation of such a state has been described in detail in \cite{gu2019quantum} (section III, Fig.4 in that manuscript). Exploiting further the technique introduced in \cite{wang2015quantum} will produce the correct ancilla state. 

Quantum states for 3-dimensional control and $d$-dimensional quantum states can also be created using \textit{Entanglement by Path Identity} \cite{krenn2017entanglement}, combined with the very recently introduced high-dimensional Bell-state measurements and high-dimensional teleportation \cite{luo2019quantum, zhang2019arbitrary}.

\end{document}